\documentclass[british,english,english,twocolumn,showpacs]{revtex4}
\usepackage{graphicx}

\makeatletter

\usepackage{color}
\usepackage{dcolumn}
\usepackage{babel}
\makeatother

\begin{document}
\title{C$_{60}$: the first one-component gel?}
\author{C. Patrick Royall}
\affiliation{School of Chemistry, University of Bristol, Bristol, BS8 1TS, UK}
\email{paddy.royall@bristol.ac.uk}
\author{Stephen R. Williams}
\affiliation{Research School of Chemistry, Australian National University, Canberra, ACT 0200}

\begin{abstract}
Until now, gels have been formed of multicomponent soft matter systems,
consisting of a solvent and one or more macromolecular or colloidal
species. Here we show that, for sufficient quench rates, the Girifalco
model of C$_{60}$ can form gels which we identify by their slow dynamics
and long-lived network structure. These gels are stable at room temperature,
at least on the simulation timescale up to 100 ns. At moderate temperatures
around 1000 K, below the bulk glass transition temperature, C$_{60}$
exhibits crystallisation and phase separation proceeds without the
dynamical arrest associated with gelation, in contrast to many colloidal systems. 
\end{abstract}
\maketitle

\section{Introduction}

Gels form part of our everyday lives, yet are remarkably poorly defined.
At moderate timescales, gels lack a clear distinction from `attractive
glasses' at high densities \cite{zaccarelli2009}, while at low densities
gels share many characteristics of liquids. Recently however, a distinction
has been made between equilibrium and non-equilibrium gels \cite{zaccarelli2007}.
The former are associated with systems whose intermolecular potential
is not spherically symmetric, for example limited-valency models lead
to equilibrium gels \cite{delGado2005} and `empty liquids' \cite{bianchi2006}.
A second class of gels are formed through arrested metastable gas-liquid
phase separation. In these `spinodal gels' the metastable liquid phase
is sufficiently dense to undergo dynamical arrest, suppressing demixing
\cite{verhaegh1997,lu2008}, an extreme case of viscoelastic phase separation \cite{tanaka2000}. 
It was recently shown that the formation
of clusters in this dense liquid phase prevents relaxation to the
underlying crystal \cite{royall2008g}. A third type of gel is formed
through diffusion-limited cluster aggregation \cite{allain1995}.
Other types of gels form in polymer solutions (spinodal gels), precipitate
from solution, for example silica, and are formed by clays such as
laponite, the nature of which remains disputed \cite{jabbariFarouji2008}.

\begin{figure}
\includegraphics[width=50mm]{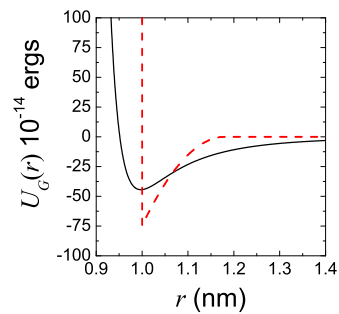}
\caption{(color online) Interaction potentials. Solid black line,
Girifalco potential used here to describe C$_{60}$ \cite{girifalco1992}.
Dashed red line, Asakura-Oosawa potential for colloid-polymer mixtures
with a size ratio of $0.18$ \cite{royall2008g}. The Asakura-Oosawa
parameters are scaled to aid comparison with the Girifalco potential.
\label{figu}}
\end{figure}

All these gels have one thing in common: they are multi-component
systems, consisting of a solvent and at least one macromolecular or
colloidal component. Although in equilibrium, the degrees of freedom
of the solvent and smaller macromolecular/colloidal species can be formally
integrated out, leading to a one-component treatment \cite{likos2001},
it is important to note that consideration of an explicit solvent
has recently been implicated in gelation, due to hydrodynamic interactions
\cite{furukawa2010}.

\begin{figure}
 \includegraphics[width=60mm]{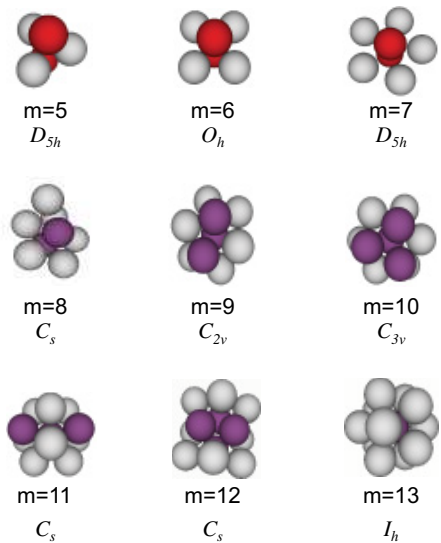}
\caption{(color online) Ground state clusters for C$_{60}$\cite{doye1996}.
Here the clusters are defined according to our topological
cluster classification. C$_{60}$ molecules are represented as spheres
\cite{williams2007}. \label{figClusters}}
\end{figure}

Here we enquire whether gels can be formed from a \emph{single} molecular
species. It has emerged from the colloid-polymer mixture literature
\cite{poon2002,zaccarelli2007} that gels are associated with systems
exhibiting a short-ranged attraction, compared to the particle/molecular
diameter. Such systems also tend not to exhibit a stable liquid phase.
Seeking a molecular system with such properties, one naturally turns
to C$_{60}$. While it is not yet clear whether C$_{60}$ has a stable
liquid phase \cite{cheng1993,hagen1993}, the commonly used model
potential of Girifalco [\ref{eqGirifalco}] \cite{girifalco1992},
is understood to exhibit a liquid over a limited range in temperature
\cite{costa2003}. The short-ranged attractions in C$_{60}$ , relative
to its molecular size, make it a most suitable candidate for one-component
`spinodal gelation'. We shall thus pose the question: can C$_{60}$
form a gel?

\begin{figure}
\includegraphics[width=60mm]{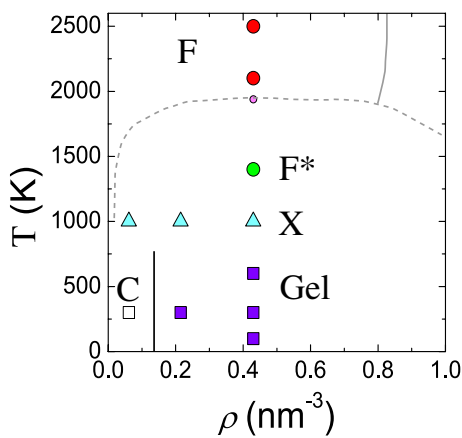}
\caption{(color online) Phase diagram for C$_{60}$. Light dashed line is
a guide to the eye denoting the liquid-gas binodal and solid line
indicates freezing (from \cite{costa2003}). Percolation at low temperature
is indicated by the solid black line.
State points sampled with constant temperature simulations and corresponding
structures are denoted as follows: (F) supercritical fluid, red circles;
(F{*}) non-phase separated metastable fluid green circle; (X) phase-separated
crystal, blue triangles; (C) isolated clusters, open square; gels,
purple squares. \label{figPhaseDiagram}}
\end{figure}

Before proceeding, let us be clear about what we mean by gelation
in this paper. Certainly, gels are percolating networks. In the case
of C$_{60}$, dynamical arrest will be required to prevent phase separation
on the measurement timescale. To create such a spinodal gel, we shall
therefore quench C$_{60}$ into the region of the phase diagram where
gas-liquid phase separation may form a bicontinuous spinodal-like
texture and undergo dynamical arrest. By dynamical arrest, we require
that the percolating structure persists on the simulation timescale.
Our requirement that phase separation is suppressed on the molecular lengthscale distinguishes
our approach from materials such as fumed silica, which is formed
from agglomerates of particles of condensed amorphous silica \cite{ihler1979}.

As in a previous study on a colloidal system \cite{royall2008g},
we shall characterise the structure in terms of clusters formed by
small groups of C$_{60}$ molecules in isolation. The structures formed
are then analysed in terms of topologies based on the minimum energy
clusters using an algorithm we have recently developed for identifying
clusters in condensed phases, the topological cluster classification
(TCC) \cite{williams2007}. Clusters of C$_{60}$ molecules have received
considerable attention \cite{baletto2005} and for small clusters
the ground states are known both experimentally \cite{martin1993,branz2000nice,branz2002,baletto2005}
and computationally \cite{doye1996}. Moreover, numerical studies have
shown that for larger clusters C$_{60}$ exhibits kinetic trapping,
leading to a discrepancy between predicted ground state structures
\cite{doye1996} and those observed experimentally below $500$ K,
which required annealing to access the ground state \cite{branz2000nice,branz2002}.

We shall use molecular dynamics simulation to capture the behaviour
of C$_{60}$ on a timescale and system size sufficient that we may
discuss gelation. For this purpose, we use the intermolecular interaction
potential proposed by Girifalco \cite{girifalco1992}, which reads

\begin{eqnarray}
u(s)=\alpha_{1}\left({\frac{1}{s(s-1)^{3}}+\frac{1}{s(s+1)^{3}}-\frac{2}{s^{4}}}\right) \nonumber \\
+\alpha_{2}\left({\frac{1}{s(s-1)^{9}}+\frac{1}{s(s+1)^{9}}-\frac{2}{s^{10}}}\right)
\label{eqGirifalco}
\end{eqnarray}

\noindent where $s=r/2a$ and $2a=7.1$ Angstroms, $\alpha_{1}=74.94\times10^{-15}$ erg,
$\alpha_{2}=135.95\times10^{-18}$ erg. The Girifalco potential is
plotted in \ref{figu}. This Girifalco model is known to undergo dynamical
arrest at high temperature and density \cite{abramo2004,greenall2006,costa2007}.
The glass line intersects the gas-liquid spinodal at around $T_{G}\approx1100$
K, while crystallisation is expected at higher temperature \cite{abramo2004}.
While locally crystalline structures still meet our criterion for
gelation (ie a long-lived network), the absence of vitrification of
the dense phase suggests that demixing may proceed. To prevent demixing,
we need to quench below $T_{G}$. Noting that free surfaces enhance
local diffusion~\cite{royall2008g,fakhraai2009,puertas2004}, we expect that it
will be necessary to quench well below $T_{G}$. This is consistent
with experimental work where annealing to $610$ K enabled small clusters
to equilibrate~\cite{branz2002}. Figure 1 also depicts the Asakura-Oosawa (AO) potential
~\cite{asakura1954,vrij1976} often used to described colloid-polymer
mixtures. The AO potential is plotted here to correspond to a 
polymer-colloid size ratio of 0.18~\cite{royall2008g} which leads
to a relative interaction range rather shorter than that of the Girifalco potential.

Our procedure is therefore as follows: we simulate Girifalco C$_{60}$
at constant temperature, following an `instantaneous' quench to enquire
at what, if any, temperature, C$_{60}$ forms a gel stable on the
simulation timescale. We then perform quenches from high temperature
to determine what quench rate is required to form gels.

\begin{figure*}
\includegraphics[width=40mm]{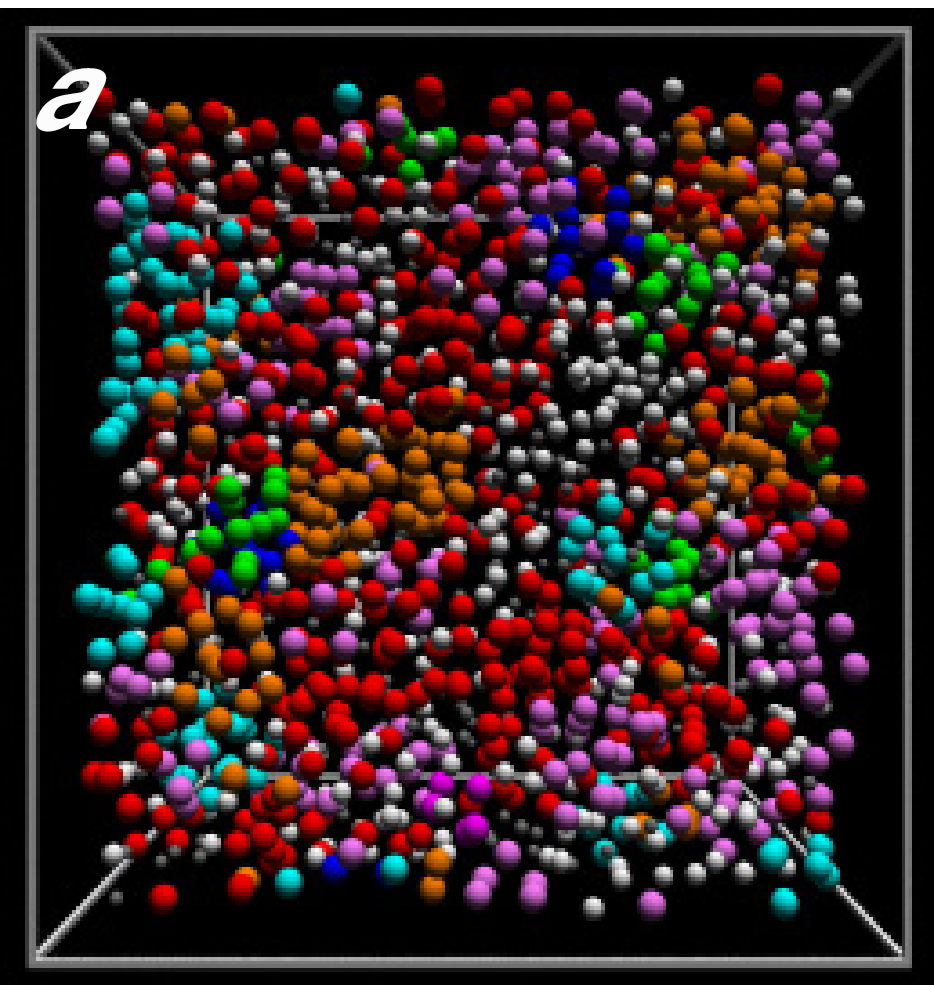}
\includegraphics[width=39mm]{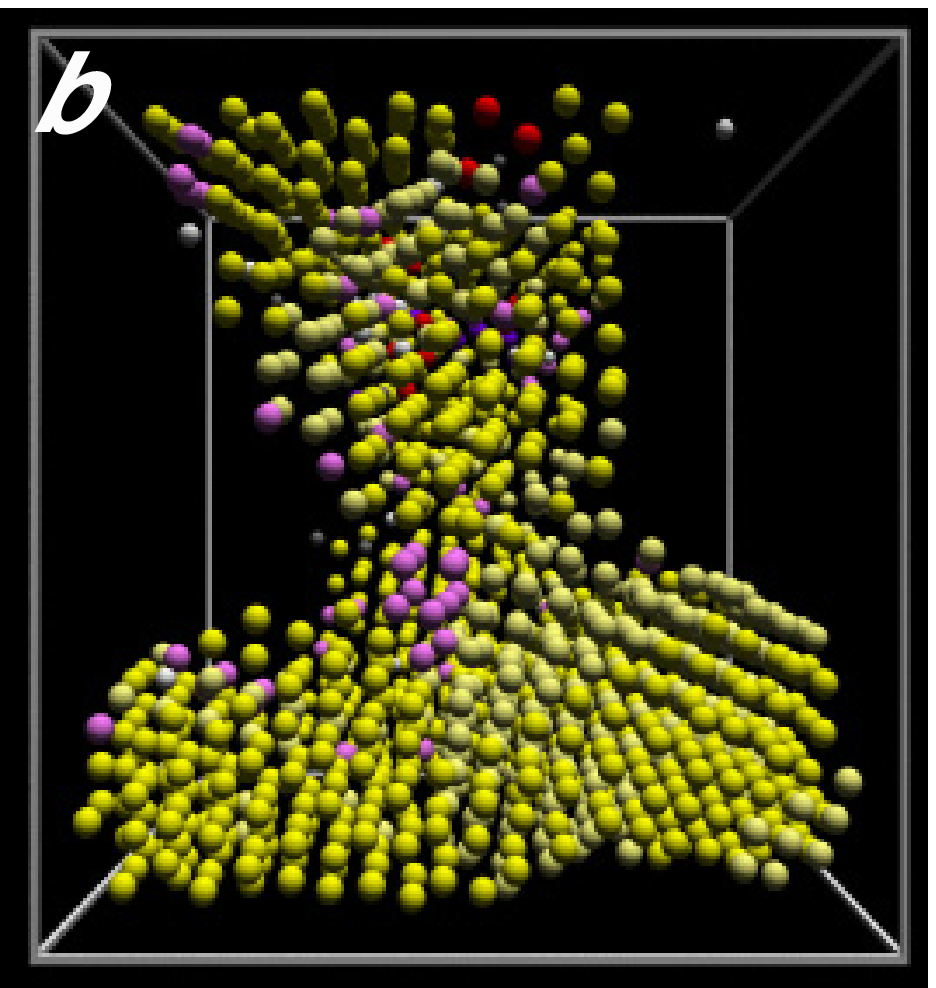}
\includegraphics[width=41mm]{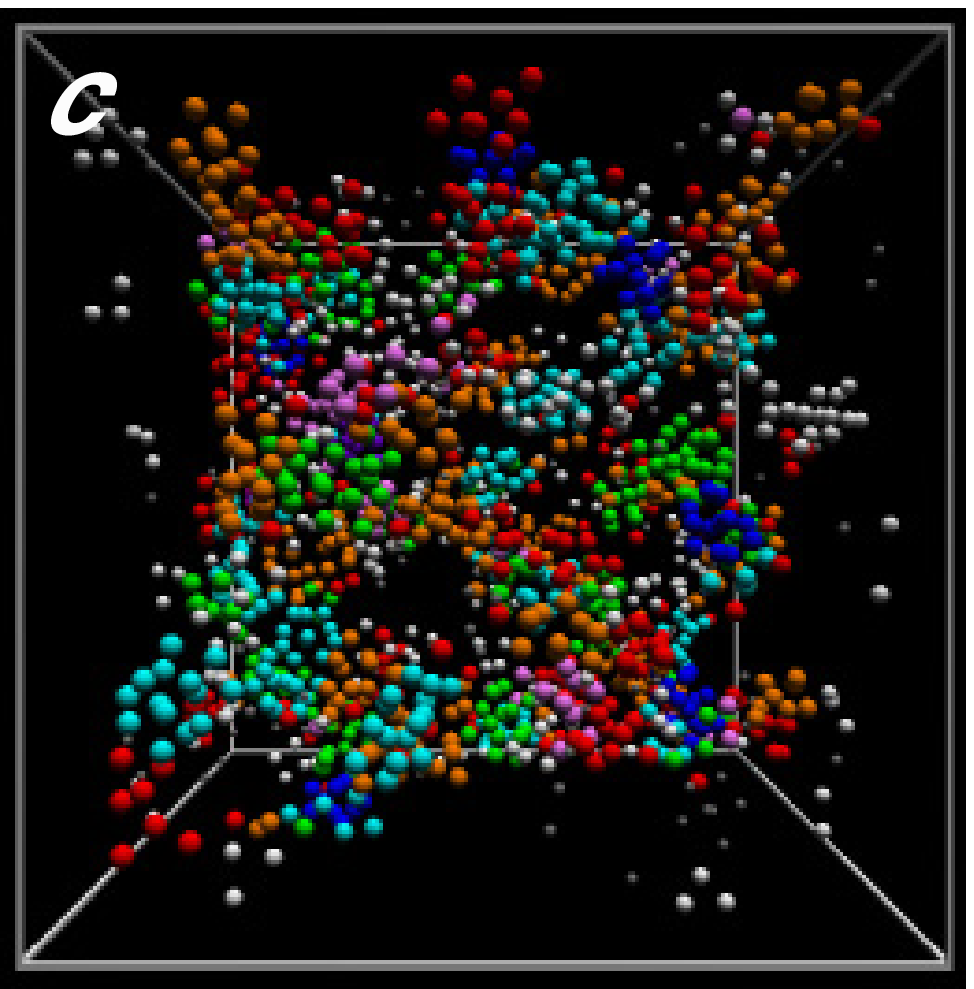}
\includegraphics[width=48mm]{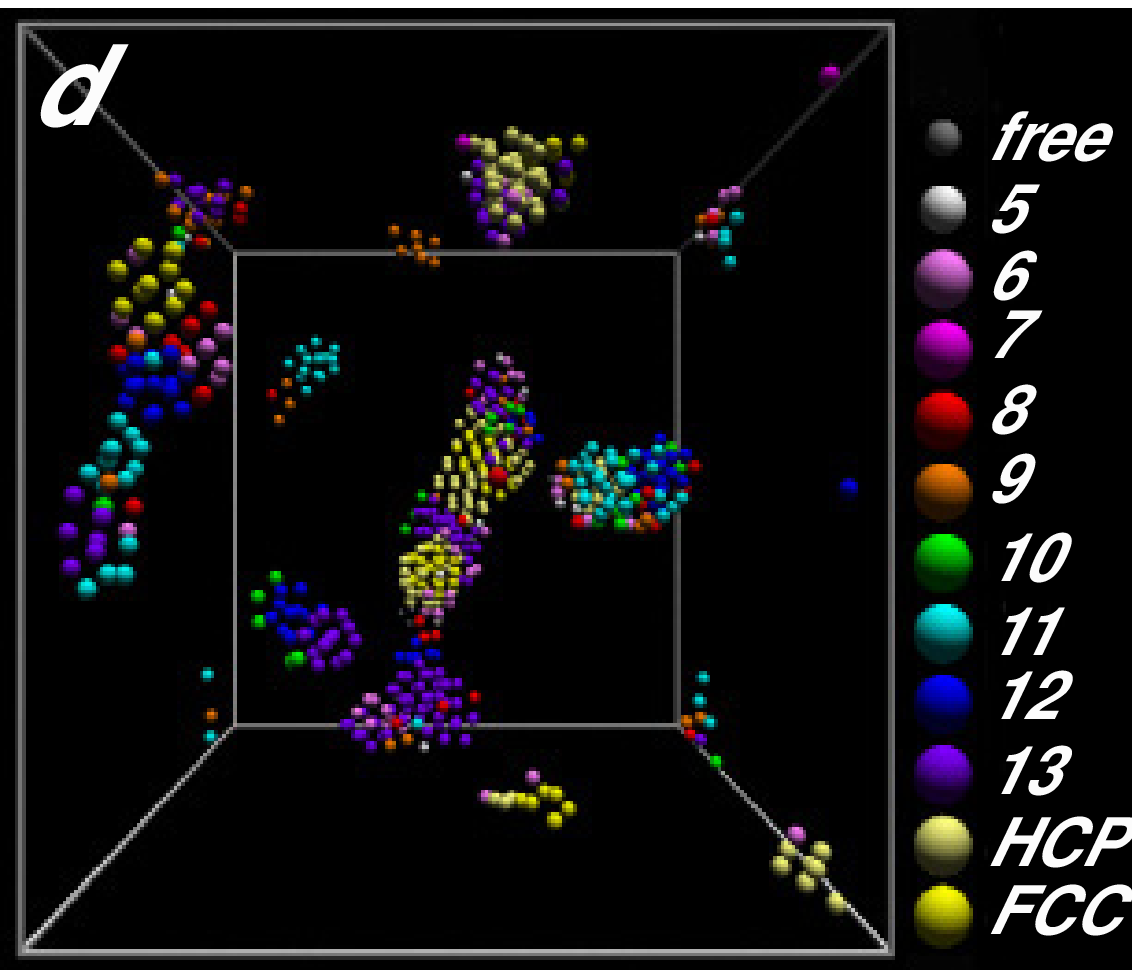}
\caption{(color online) Typical (non-equilibrium) structures. (a) non-phase separated fluid ( $\rho=0.43$ nm$^{-3}$, $T=1400$ K), (b) phase separated crystal ($\rho=0.430$ nm$^{-3}$, $T=1000$ K), (c) gel ( $\rho=0.215$ nm$^{-3}$, $T=300$ K); (d) isolated clusters ($\rho=0.108$ nm$^{-3}$, $T=300$ K). All data shown here are from simulations at constant temperature. Colors denote the differing clusters or crystal structures to which the molecules belong, as shown on the right. Free denotes molecules not belonging to any clusters. Molecules are drawn around half actual size and represented as spheres. \label{figPretty} }
\end{figure*}

\section{Simulation details}

\subsection{Molecular dynamics simulation}

We carry out simulations of two types. Constant temperature simulations,
where the system is run up to $10^{7}$ timesteps ($99.6$ ns) and sampled
for a further $10^{6}$ timesteps, and temperature quenches where
temperature is decreased from $2328$ $\mbox{K}$ to $300$ $\mbox{K}$
at rates between $2.04\times10^{11}$ and $2.04\times10^{13}$ Ks$^{-1}$.
State points sampled for constant temperature simulations are shown
in \ref{figPhaseDiagram}. Throughout we use molecular dynamics
simulation with a Gaussian thermostat \cite{evans2008} with a timestep
of $9.96$ fs. We typically used 4000 or 2048 C$_{60}$ molecules. Runs with
4000 particles showed no qualitative differences. Occasionally, long
times required smaller simulations of 864 particles. Data are shown
after 9.96 ns of `equilibration' or `aging', unless otherwise stated.
All simulations are performed at constant volume. We focus on data
taken along the critical isochore $\rho\approx0.43$ nm$^{-3}$ unless 
otherwise stated. Results at lower density $\rho\approx0.215$ nm$^{-3}$ 
are qualitatively similar.

\subsection{The topological cluster classification}

To analyse the structure, we identify the bond network. Here we define
two molecules as bonded if they approach within $1.23$ nm, which
is the approximate location of the first minimum of the pair correlation
function $g(r)$ upon condensation. Having identified the bond network,
we use the Topological Cluster Classification (TCC) to determine the
nature of the cluster~\cite{williams2007}. This analysis identifies
all the shortest path three, four and five membered rings in the bond
network. 
We use the TCC to identify structures of between 5 and 13 molecules 
which are topologically equivalent to ground states for the Girifalco 
potential (\ref{eqGirifalco}) \cite{doye1996}.
The clusters we identify are depicted in \ref{figClusters}.
In addition we identify the FCC and HCP thirteen particle structures
in terms of a central particle and its twelve nearest neighbours.
For spherically symmetric potentials differing cluster structures
are found as a function of interaction range \cite{doye1995}. For
11-membered clusters, our topological approach does not distinguish
between $C_{s}$ (11D in \cite{doye1995}) and $C_{2v}$ (11C in \cite{doye1995}).
Likewise for 12-membered clusters $C_{s}$ (12C in \cite{doye1995})
and $C_{5v}$ (12B in \cite{doye1995}) are identical for our purposes.
For more details see \cite{williams2007}. If a molecule is found
to be part of more than one cluster, we count it as the larger cluster.
Moreover, if a molecule is found to be part of both a HCP and FCC
local environment (for example in random close packed stacking), we
count it as FCC.

\section{Results}

\subsection{Constant temperature simulations}

We begin our presentation of the results by discussing our interpretation
of the different structures we find. The equilibrium phase diagram
is sketched in \ref{figPhaseDiagram} \cite{costa2003} and typical
structures are shown in \ref{figPretty}. However, the system
is not always able to fully equilibrate on the simulation timescale, 
a point to which we return below.
For the densities we consider, at temperatures less than the triple
point (around 1880 K \cite{costa2003}), equilibrium for the Girifalco
model of C$_{60}$ is gas-crystal coexistence. However, for weak degrees
of undercooling, crystal nucleation may not occur on the system sizes
and timescales we consider, while for deep quenches, dynamical arrest
limits access to equilibrium. In our simulations, these two non-equilibrium
states are manifested as a metastable fluid, which can (but does not
always) undergo phase separation prior to crystallisation, and a gel
which forms a percolating network respectively. We identify
these states by visual inspection of coordinate data and also require
that more that 50\% of molecules are in a locally HCP or FCC environment
according to the TCC for any state point to be considered to be crystalline.
We further require that gel states exhibit slow dynamics, in other words,
that they do not phase separate or significantly coarsen on the simulation timescale.

More rigorous approaches, notably testing for percolation, have the
disadvantage that for the system sizes we are able to tackle here,
for $\rho\gtrsim0.2$ nm$^{-3}$, all state points were found to percolate
according to our bonding criterion, despite the different structures
evident in \ref{figPretty}. At low density, isolated clusters
were found [\ref{figPretty}(d)]. In our simulations, these
clearly do not form a percolating network. Thus we sketch a percolation
line in \ref{figPhaseDiagram}.

\begin{figure}
 \includegraphics[width=40mm]{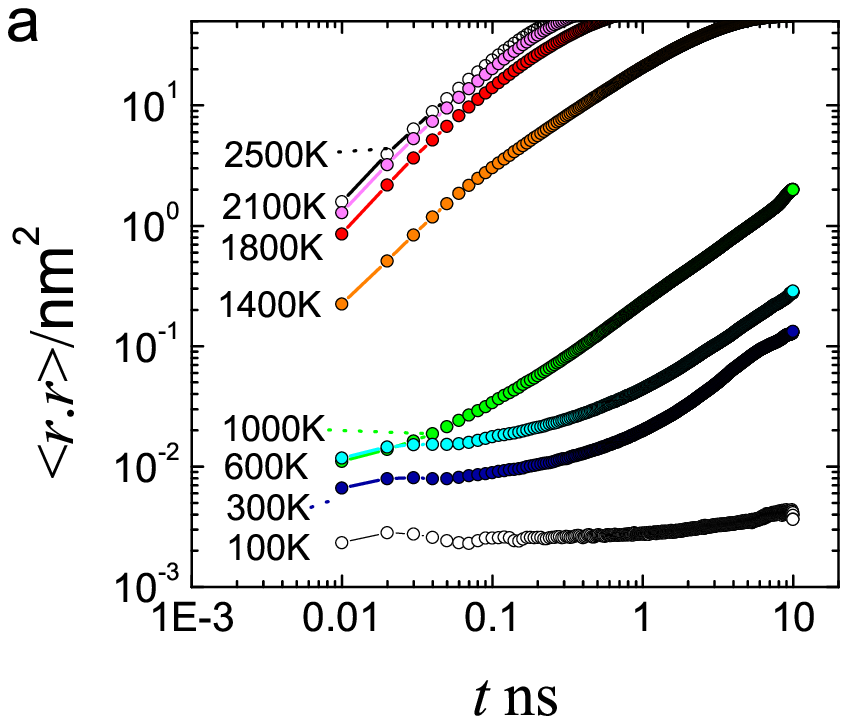}
 \includegraphics[width=45mm]{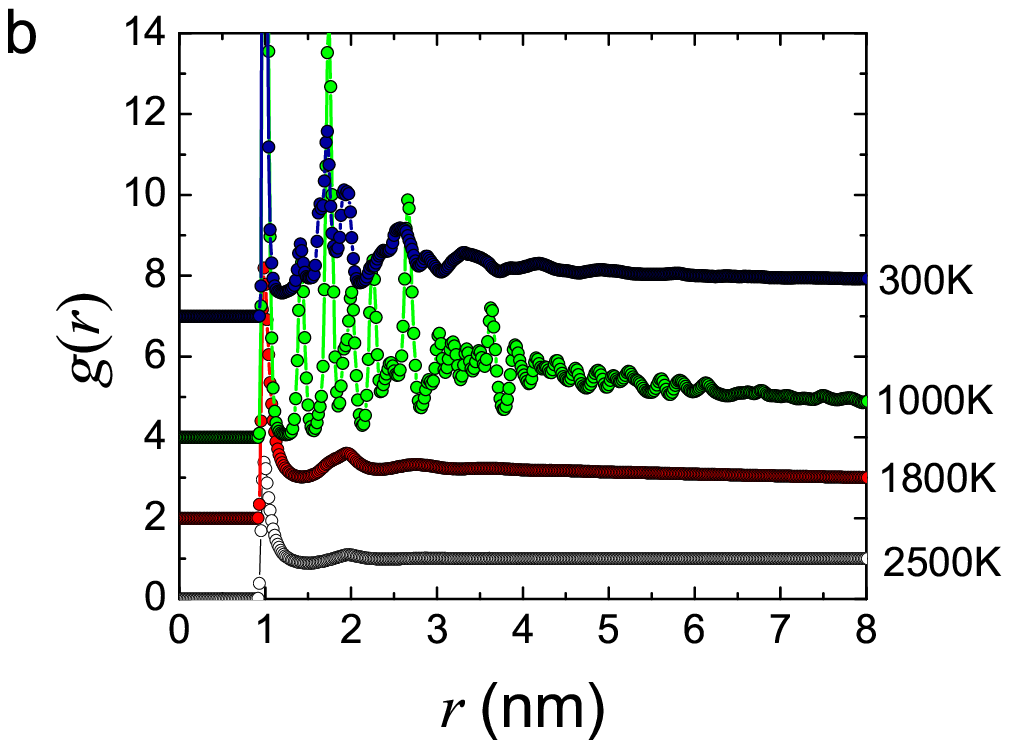}
\caption{(color online) Mean squared displacement (a) and pair correlation functions (b) for constant temperature simulations. Here $\rho=0.43$} nm $^{-3}$. In (b) the data are offset for clarity. \label{figMSDg}
\end{figure}

We have argued that in the case of C$_{60}$, gelation should be associated with slow dynamics.
Plotting the mean squared displacement (MSD) at different temperatures
in \ref{figMSDg}(a), we see a sharp fall in MSD between T=1400
and 1000 K. Thus we expect that, at 1000 K and below, we may find
gelation. Inspection of the coordinate data in \ref{figPretty}(b)
shows that at 1000 K, C$_{60}$ crystallises on our simulation timescales.
This is consistent with nucleation studies carried out at higher density
\cite{ngale2009}. However at some of the lower
temperatures we consider, it is clear C$_{60}$ is sufficiently arrested that little
restructuring is seen on the simulation timescale. Dynamically, therefore,
it is possible that C$_{60}$ can satisfy our requirements for classification
as a gel. We note that these measures are performed on a system
which is not in equilibrium for some state points. However, our intention
here is to \emph{qualitatively} demonstrate a substantial slowing of the dynamics.

\begin{figure}
 \includegraphics[width=65mm]{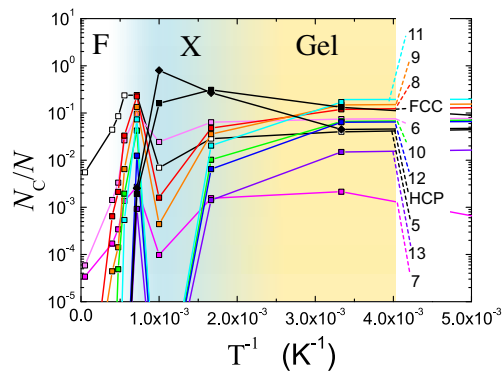}
\caption{(color online) The population of molecules in ground state clusters
of C$_{60}$ as a function of inverse temperature. Here $\rho=0.43$
nm$^{-3}$. Shaded regions are a guide to the eye. Note the
semi-log scale. \label{figTCC}}
\end{figure}

Turning to local structural measures, namely the pair correlation
function $g(r)$ we see three regimes in \ref{figMSDg}(b). At
high temperatures, a fluid $g(r)$ is found. Note that even at temperatures
below the triple point \cite{costa2003}, we see little change in
$g(r)$ on these simulation timescales. However, at lower temperatures (1000 K),
the system crystallises. At lower temperatures still (300 K), although
there is some local structure, overall the ordering is reduced. For
temperatures less than around 1000 K $g(r)$ takes values
substantially greater than unity at short range. This indicates density fluctuations
on the lengthscale of a few particle diameters. While the limited
size of our simulations preclude a definitive statement, we note that
such local density fluctuations are indicative of networks, such a that found 
in \ref{figPretty}(c).

Using the topological cluster classification, the same three regimes
are identified, as shown in \ref{figTCC}. At high temperatures
(T$\gtrsim$1400 K), the population of all clusters rises sharply
upon reducing the temperature.
Around 1000K, the structure is dominated by molecules in a locally
crystalline environment, and the population of other clusters drops.
At lower temperatures still, the population of locally FCC and HCP
molecules drops, and the system is once again dominated by amorphous clusters.
Of these, at 300 K, 11-membered $C_{s}$ clusters are the most numerous,
accounting for some 20\% of the total population of C$_{60}$ molecules.
These clusters are five-fold symmetric, and it is tempting to infer
that this five-fold symmetry might impede crystallisation \cite{frank1952}.
The five-fold symmetric structure popularised by Frank \cite{frank1952}, the 13-membered
icosahedron is less popular than either HCP or FCC in this case, despite
being the minimum energy cluster for 13 molecules \cite{doye1996}. We note that the
icosahedron has been found to be relatively resistant to kinetic trapping
\cite{baletto2002} and is also observed experimentally \cite{martin1993}.

\begin{figure}
\protect\includegraphics[width=42mm]{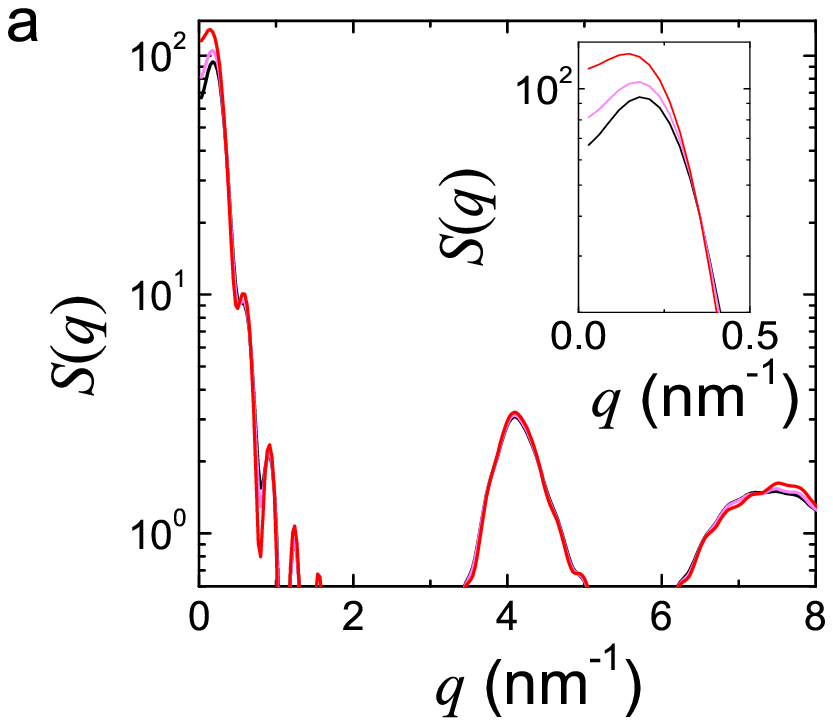} 
\protect\includegraphics[width=41mm]{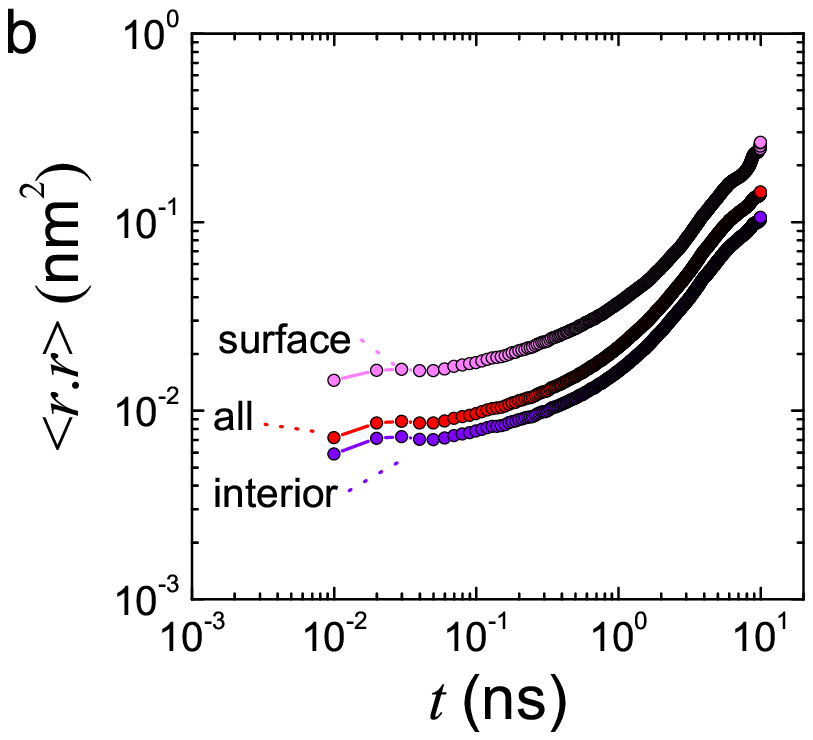}
\caption{(color online) (a) Static structure factor $S(q)$ for gels at different ageing times $t_w$. Black line $t_w=1$ ns, pink line, $t_w=10$ ns, red line, $t_w=100$ ns. Inset shows first peak. $\rho=0.21$} nm\textbf{$^{-3}$} $T=300$ K. (b) Mean squared displacement for surface (pink), and interior (purple) molecules. $\rho=0.43$ nm\textbf{$^{-3}$ $T=300$ K.
\label{figStability}}
\end{figure}

We now consider the gel stability. Spinodal gels such as those we consider here
are metastable to phase separation. Although
that process is suppressed by slow dynamics induced by quenching, it is reasonable to pose the question, to what
extent is phase separation effectively suppressed? To this end, we plot the static structure factor $S(q)$
in Fig. \ref{figStability}(a). Indeed a small degree of structural evolution is found, as shown by the 
slight shift in the first peak of $S(q)$. While this indicates some degree of coarsening, it is very clear that the
gel lifetime exceeds the simulation timescale, noting that here the simulations are run for up to 100 ns. 

Nevertheless, there is some degree of coarsening. We investigate possible coarsening mechanisms by 
considering the mean squared displacement of particles on the surface and in the interior of the `arms' of the gel. We define molecules on the surface as those with fewer than eight neighbours. 
The analysis in Fig. \ref{figStability}(b) shows that indeed, molecules on the surface have a marginally higher mobility
than those in the interior. This is qualitatively consistent with previous studies of gelation in colloidal systems~\cite{royall2008g,puertas2004}.

\begin{figure}
\protect\includegraphics[width=39mm]{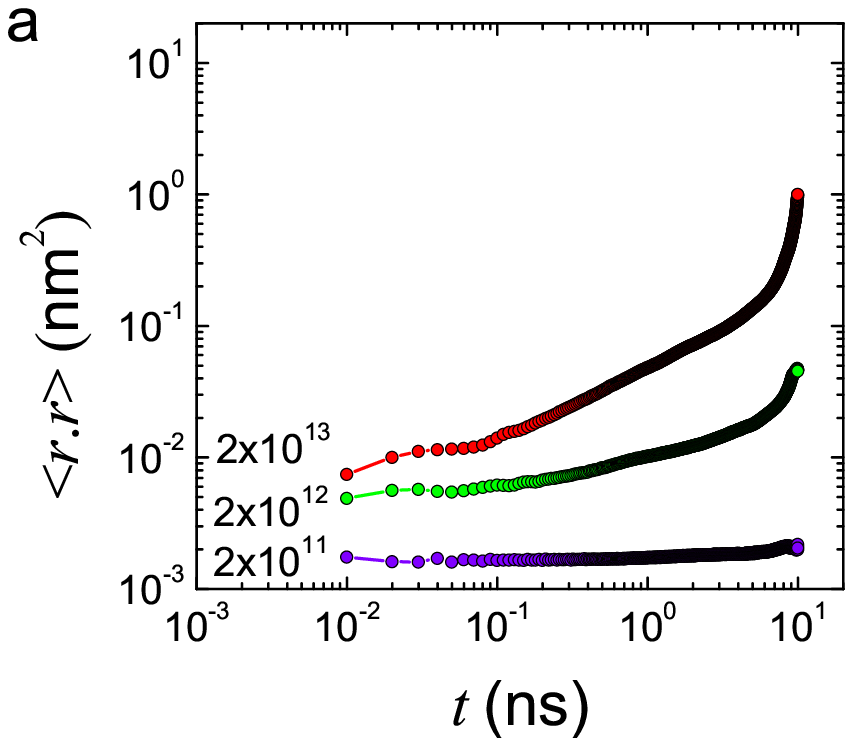}
\protect\includegraphics[width=45mm]{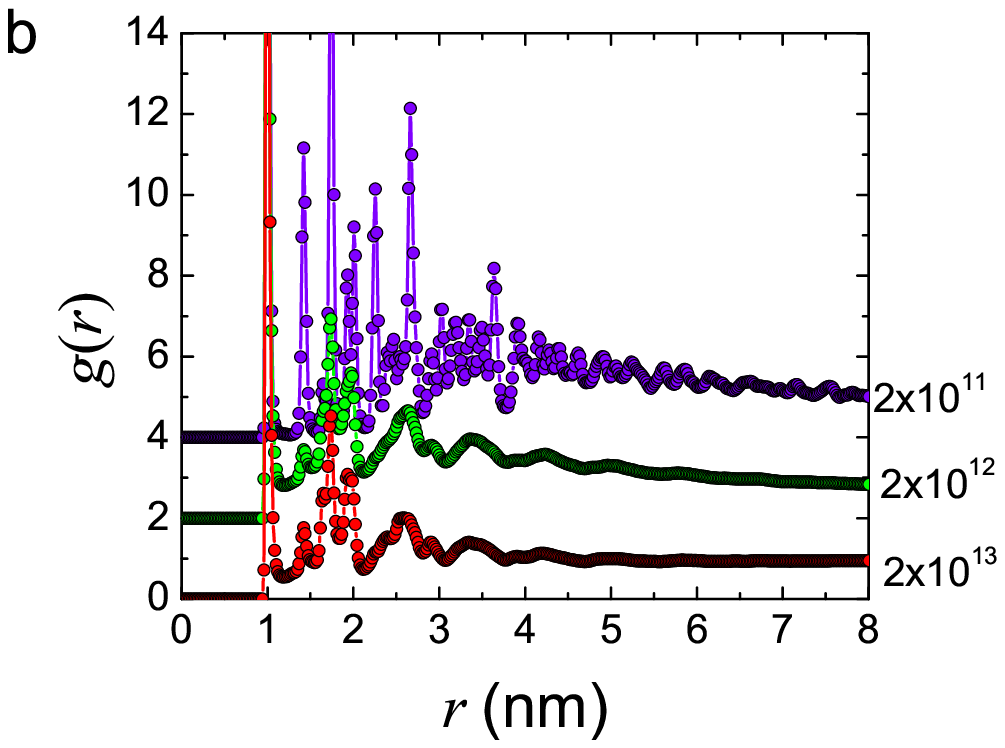} 
\caption{(color online) Mean squared displacement (a) and pair correlation functions (b) for quenches. Here $\rho=0.43$} nm$^{-3}$. Quench rate is denoted in Ks$^{-1}$.\label{figMSDgQuench}
\end{figure}

\subsection{Temperature quenches}

The picture that emerges from our analysis of the constant temperature
runs is that at T$\lesssim$$600$ K, C$_{60}$ can form gels, according
to our criteria. However such `instantaneous quenches' are unphysical
for molecular systems (although essentially accessible to soft matter
systems such as colloids). We therefore enquire as to what quench rate
is required to form a gel. We choose to quench from a supercritical
fluid at $2328$ K to $300$ K. Inspection of the phase diagram suggests
that, for sufficiently slow quench rates, we might expect phase separation
to occur in temperature regime $1940$ K $\gtrsim$ T$\gtrsim$$600$
K, and that crystallisation may also occur.

We thus consider a range of quench rates, and identify the final state
as before. Data are presented for the $9.96$ ns immediately following
the quench. Mean squared displacement data [\ref{figMSDgQuench}(a)]
shows that the quenches we have performed result in a state with relatively
slow dynamics. Mobility is strongly suppressed for slower quench rates.
The pair correlation function data in \ref{figMSDgQuench}(b)
show a similar structural transitions to the constant temperature
data, although here the transition occurs as a function of quench
rate. For quench rates at or above $2.04\times10^{12}$ Ks$^{-1}$,
we find that the $g(r)$ remains amorphous. A slower quench rate of
$2.04\times10^{11}$ Ks$^{-1}$results in a crystalline structure,
which undergoes phase separation, similar to \ref{figPretty}(b).

\begin{figure}
 \includegraphics[width=69mm]{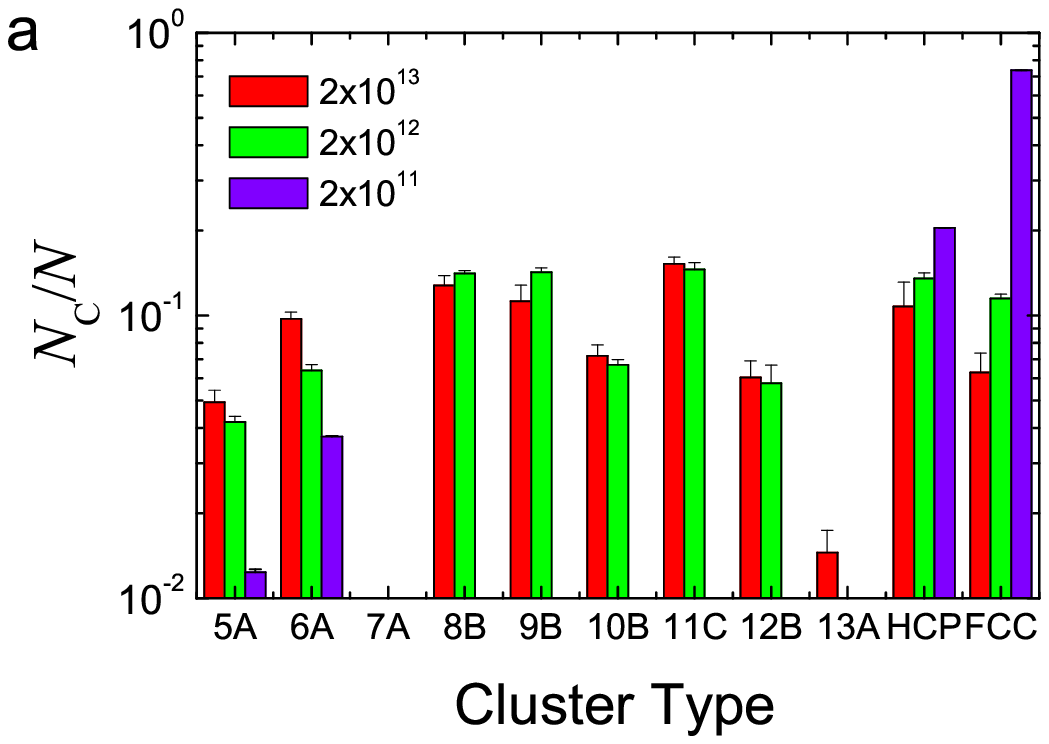}
 \includegraphics[width=70mm]{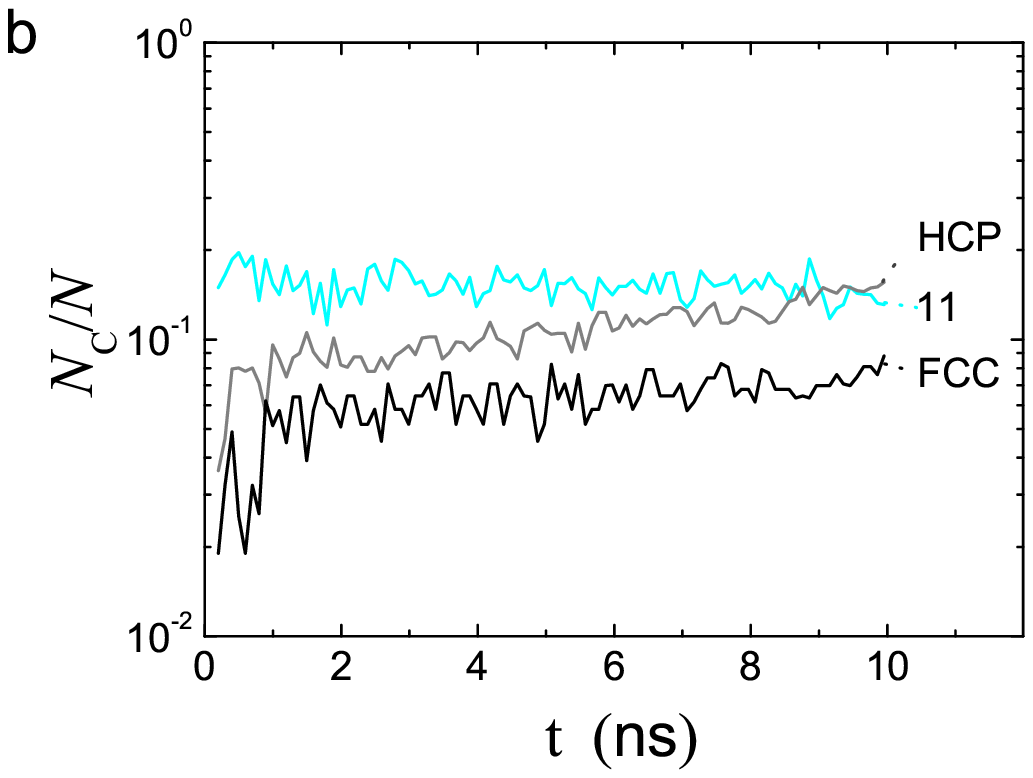}
\caption{(color online) The population of molecules in ground state clusters of C$_{60}$ for different quench rates (a). Quench rates are denoted in Ks$^{-1}$. (b) Structural changes upon aging in a gel formed after a quench rate of $2.04\times10^{13}$Ks$^{-1}$. Here $\rho=0.43$ nm$^{-3}$. \label{figQuenchTCC} }
\end{figure}

This observation is confirmed with the TCC analysis, where we consider
the cluster populations in the quenched systems. In \ref{figQuenchTCC}(a)
we show average cluster populations for runs of 9.96 ns following
the quench, and find a clear shift to a crystal-dominated structure
for the slowest quench rate. In fact the trend towards a higher degree
of crystalinity is apparent between the higher quench rates of $2.04\times10^{13}$
and $2.04\times10^{12}$ Ks$^{-1}$. Turning to the mean squared displacement
data in \ref{figMSDgQuench}(a) we note that for rapid quenching
in particular, some movement persists. In other words, the gel undergoes
aging. Restructuring associated with this aging is apparent in the larger error bars for the $2.04\times10^{13}$
Ks$^{-1}$quench rate in \ref{figQuenchTCC}(a). In \ref{figQuenchTCC}(b)
we consider the local structural consequences of this aging. The data
shows a gradual trend of increasing HCP and FCC local structures and
a decrease of the most popular cluster, $m=11$ $C_{2v}$. This increase in crystalinity
is consistent with previous studies on aging colloidal gels \cite{darjuzon2003}.

\section{Discussion and Conclusions}

We begin the discussion by considering how practically realisable
the C$_{60}$ gels we have predicted might be. We have shown that gels
form and exist on a simulation timescale of 10 ns. This leads to two
main questions: firstly, can the gels form in the first place and
secondly, how long can they last? Evidence from experimental work
on clusters of C$_{60}$ molecules suggests that in fact, long-lived
metastable states are possible. Below around 500 K, clusters show
little preference for ground states \cite{branz2000nice,branz2002,baletto2005}.
We thus argue that C$_{60}$ can be prevented from equilibrating on
an experimental timescale at room temperature, although we note that
some aging can be seen in our simulations. The first question is potentially
more challenging. The quench rates we have used are not currently
experimentally feasible (maximum quench rates are around 10$^{6}$
Ks$^{-1}$by vapor deposition). Our predictions are that C$_{60}$
will phase separate to crystal-gas coexistence on such a timescale.
However, shear could be used to prevent crystallisation during quenching,
allowing milder quench rates. Other possibilities include compressing
low-temperature amorphous clusters. We note in \ref{figPretty}(d)
that the isolated clusters formed at T$=300$ K are far from spherical
and might reasonably be expected to form a gel upon increasing the
density. Following \cite{baletto2002} we also note that the Girifalco
potential is not the only model for C$_{60}$. The Pacheco-Prates-Ramalho 
model \cite{pacheco1997} is in fact more sticky (shorter ranged) than that of Girifalco, which
would be expected to promote gelation.

Another approach to producing one-component gels might be to consider
larger fullerenes whose dynamics will be much slower. Such a candidate
is C$_{540}$which may also exhibit a molecular shape with high symmetry
and stability \cite{scuseria1995}, although its high binding energy
would necessitate quenching from very high temperatures at which thermal
stability of the fullerene molecule may become a problem. The same
might even hold for C$_{60}$ whose thermal stability at high temperatures
(such as criticality) is disputed \cite{kolodney1994,sommer1996},
although at temperatures below 1700 K C$_{60}$ is likely stable,
so compressing disordered clusters remains a potential experimental route
to forming gels of C$_{60}$. 

The behaviour exhibited by C$_{60}$ in the constant temperature simulations
may be compared with that observed in colloidal suspensions, where
interactions may be tuned \cite{poon2002}, and which typically undergo
`instantaneous' quenching, due to the slow relaxation time of the
mesoscopic colloidal particles. The main difference between the results
presented here and those from a previous study on a colloidal gel
\cite{royall2008g} are that C$_{60}$ exhibits a regime of crystallisation,
which is almost entirely absent from the colloidal system, where the crystal-like
local structures accounted for no more than one particle in 1000.
There are at least three possible reasons for this difference. The
first is that the relative interaction range is substantially longer
in the case of C$_{60}$ and this may facilitate crystallisation (see
\ref{figu}). Secondly, colloids are polydisperse, which might
suppress, although should not prevent, crystallisation \cite{moriguchi1995,williams2001}.
Finally, we have used Newtonian dynamics for C$_{60}$ while colloids
are diffusive, although we are unaware of any studies considering
the role of dynamics in crystallisation. In colloid simulation work with 
Brownian dynamics~\cite{fortini2008}
with parameters comparable to experiments~\cite{royall2008g}  
there is some evidence that crystallisation can occur 
on the low temperature side of the metastable gas-liquid binodal in a monodisperse system. This 
suggests that the role of polydispersity in the colloid experiments
may be important in suppressing crystallisation.

In summary, we have presented numerical evidence that C$_{60}$, under
the right conditions can form a gel, which forms through arrested
spinodal decomposition. Potentially, therefore, one-component gels
might be realised. Whether these conditions are experimentally accessible
remains an open question.

\begin{acknowledgements} CPR acknowledges the Royal Society for financial
support. The authors would like to thank Hajime Tanaka for many helpful
discussions on the nature of the gel state, with particular reference
to Laponite. 
\end{acknowledgements}


\providecommand*\mcitethebibliography{\thebibliography}
\csname @ifundefined\endcsname{endmcitethebibliography}
  {\let\endmcitethebibliography\endthebibliography}{}

\end{document}